\documentclass[12pt]{article}

\usepackage{newtxtext,newtxmath}
\usepackage[T1]{fontenc}
\usepackage{graphicx}
\usepackage{soul}

\usepackage[letterpaper,margin=1in]{geometry}

\renewenvironment{abstract}
	{\quotation}
	{\endquotation}

\date{}

\makeatletter
\renewcommand{\fnum@figure}{\textbf{Figure \thefigure}}
\renewcommand{\fnum@table}{\textbf{Table \thetable}}
\makeatother

\usepackage{scicite}

\usepackage{url}

\renewcommand{\color}[1]{} 

\def\scititle{
	Nanoscale Polar Landscapes in Quantum Paraelectric SrTiO$_{3}$
}
\title{\bfseries \boldmath \scititle}

\author{
	Yang Zhang$^{1,5\ast}$,
    Suk Hyun Sung$^{1,7}$,
    Nishkarsh Agarwal$^{2}$,
    Maya Gates$^{3}$,\and
    Cong Li$^{4}$,
    Pu Yu$^{4}$,
    Robert Hovden$^{2,8}$,
	Ismail El Baggari$^{1,6\ast}$
	\and
	\small$^{1}$The Rowland Institute at Harvard, Harvard University, Cambridge, MA, 02138, USA.\and
	\small$^{2}$Department of Materials Science and Engineering, University of Michigan, Ann Arbor, MI, 48105, USA\and
    \small$^{3}$h-Bar Instruments, Ann Arbor, MI, 48103, USA\and
    \small$^{4}$State Key Laboratory of Low Dimensional Quantum Physics and Department of Physics, \and \small Tsinghua University, Beijing, 100084, China.\and
    \small$^{5}$Department of Mechanical Engineering and Materials Science Duke University, Durham, NC, 27708, USA.\and
    \small$^{6}$Department of Physics $\&$ Astronomy, University of British Columbia, Vancouver, BC, V6T1Z4, Canada.\and
    \small$^{7}$Michigan Center for Materials Characterization, University of Michigan, Ann Arbor, MI, 48105, USA.\and
    \small$^{8}$Applied Physics Program, University of Michigan, Ann Arbor, MI, 48105, USA\and
	\small$^\ast$Corresponding author. Email: yang.zhang@duke.edu, ismail.elbaggari@ubc.ca
}

\begin{document} 

\maketitle

\begin{abstract} \bfseries \boldmath
SrTiO$_{3}$ is a textbook quantum paraelectric, with ferroelectricity purportedly suppressed by quantum fluctuations of ionic positions down to the lowest temperatures.   
The precise real space structure of SrTiO$_{3}$ at low temperature, however, has remained undefined despite decades of study. 
Here we directly image the low-temperature polar structure in the quantum paraelectric phase of a SrTiO$_{3}$ lamella, using cryogenic scanning transmission electron microscopy down to 20 K. 
High resolution imaging reveals a spatially fluctuating landscape of nanoscale domains of finite polarization. 
The short-range polar domains first grow and self-organize into a periodic structure over tens of nanometers. 
However, the process reverses when entering the quantum paraelectric regime below 40 K and the periodically ordered polar nanodomains fragment into small clusters.
\end{abstract}

When cooling toward a ferroelectric phase transition, collective atomic motions (phonons) slow down (soften), until a static atomic displacement pattern sets in—a spontaneous polarization appears throughout the ferroelectric material \cite{cochran1960crystal}. 
However, in quantum paraelectrics, ferroelectric order does not set in at low temperature: 
persistent quantum fluctuations of the ionic positions preclude long range ordering of polarization \cite{muller1979srti}. 
The perovskite strontium titanate (SrTiO$_{3}$) is the prototypical quantum paraelectric \cite{muller1979srti, martovnak1994path,rowley2014ferroelectric}. 
Quantum fluctuations in SrTiO$_{3}$ are associated with a range of remarkable properties, including large dielectric permitivity \cite{sakudo1971dielectric, neville1972permittivity}, strain-induced ferroelectricity \cite{haeni2004room,
lee2015emergence}, multiferroicity \cite{basini2024terahertz}, and superconductivity \cite{schooley1964superconductivity, koonce1967superconducting,
collignon2019metallicity}.

Quantum paraelectricity in SrTiO$_{3}$ manifests as a large rise in the dielectric function \cite{shirane1969lattice, weaver1959dielectric, yamanaka2000evidence},
and a deviation from classical Curie-Weiss behavior below $T_{q} \approx$ 40 K.
SrTiO$_{3}$ exhibits additional anomalous dynamics that precede and influence this quantum regime. 
A transverse acoustic phonon mode partially softens at a finite wavevector, $q$, hinting to a possibly novel state with spatial correlations at the nanometer scale \cite{salje2013domains,frenkel2017imaging,fauque2022mesoscopic,muller1991indication,orenstein2025observation,guzman2023lamellar}.
Despite the tunability and widespread applications of SrTiO$_{3}$, the low-temperature phases---from which the emergent dielectric, ferroelectric, and superconducting phases are derived---remain undefined. 
Both above and below $T_{q}$, the real space structures of the low-temperature phases of SrTiO$_{3}$ have not been resolved.
Pinpointing the precise structure could unveil the nature of quantum paraelectricity in this prototypical system

Here we show that the quantum paraelectric regime in SrTiO$_{3}$ is governed by nanoscale polar domains (Fig. 1c).
Leveraging liquid helium cryogenic scanning transmission electron microscopy (STEM), we directly image the local polarity in a SrTiO$_{3}$ lamella down to very low temperatures ($\sim$ 20 K) and reveal a spatially fluctuating and evolving landscape of nanoscale polar textures. 
Polarity emerges below T$_{AFD}$, but in the form of spatially fluctuating short-range nanodomains. 
The short-range polar domains grow and, remarkably, self-organize into a periodic structure extending over tens of nanometers. 
However, below T$_{q}$, this periodic spatial ordering weakens and the polar nanodomains fragment into small and disordered clusters upon entering the quantum paraelectric regime.

{\subsection*{Low-temperature structural transitions in SrTiO$_{3}$ lamellae}}
Bulk SrTiO$_{3}$ hosts distinct structural phases (Fig. 1a). 
In the high-temperature cubic phase, the titanium ions sit at the center of the oxygen octahedron and strontium corner atoms. 
A so-called antiferrodistortive (AFD) structural transition occurs below $T_{AFD} \approx 105 \ K$ and involves rotations of oxygen octahedra in opposite directions \cite{fleury1968soft}, leading to a tetragonal phase and a "doubling" of the unit cell along the rotation axis.
Further cooling is accompanied by large rise in the dielectric function, indicating proximity to a polar, ferroelectric transition \cite{shirane1969lattice, weaver1959dielectric, yamanaka2000evidence}. 
However, in pure SrTiO$_{3}$, long-range ordering of polar displacements does not set in---purportedly due to quantum fluctuations of opposite polarization states.

The quantum paraelectric regime in SrTiO$_{3}$ lies in proximity to long-range ferroelectricity which can be accessed under large enough external tuning parameters, including strain, oxygen vacancies or isotope substitution (Fig. 1b) \cite{jang2010ferroelectricity, leroy2025antiferrodistortive, haeni2004room, li2006phase, aschauer2014competition, lee2015emergence,xu2020strain, salmani2020order,mitsui1961dielectric,itoh1999ferroelectricity,li2019terahertz,nova2019metastable,basini2024terahertz, fechner2024quenched}. 
Quantum paraelectricity is thus confined within finite window in the phase diagram, before the onset of long-range ferroelectricity.
We verify our sample lamella prepared for STEM imaging (inset of Fig. 2a) retains the key transitions associated with parent SrTiO$_{3}$.
The potential influence, including thermal mismatch strain and oxygen deficiency, is examined and excluded (Figs. S1-S2 and note I).
Figure 1d shows the emergence of AFD superspots at half-integer positions in electron diffraction patterns (marked by the black arrow) due to the out-of-phase rotation of TiO$_{6}$, which doubles unit cell along rotation axis.
The superspot intensity, initially low at elevated temperatures (Fig. 1d, complete datasets in Fig. S3), reveals a pronounced increase near $T_{AFD}$, confirming entry into the AFD state.
Rather than appearing suddenly at $T_{AFD}$, weak superspot intensity persists above $T_{AFD}$, in agreement with the documented precursor effects and diffuse intensity reported in neutron and X-ray measurements \cite{riste1971critical, shapiro1972critical, darlington1976central}.
In addition to the AFD order, the lattice parameter along the AFD rotation axis elongates upon crossing $T_{AFD}$ (Fig. 1e), in agreement with the symmetry change into a tetragonal phase expected for the bulk AFD transition.

At lower temperatures, the quantum paraelectric phase shows additional anomalies in these parameters;  
the AFD superspot intensity exhibits a drop below $T_{q}$, and the lattice constant along the AFD rotation axis contracts. 
This behavior represents independent evidence of the elastic anomaly associated with the quantum paraelectric phase \cite{tsunekawa1984linear, neumann1995investigation, kityk2000low}.
These experimental signatures indicate the lamellar sample qualitatively retains the structural transitions at $T_{q}$ and $T_{AFD}$ (note II);
however, we cannot rule out slight few-degree changes in the transition temperatures due to the coarse temperature sampling.

{\subsection*{Visualizing polarization in cryogenic SrTiO$_{3}$}}
Beyond AFD and lattice order, maps of polarization directly reveal a complex local landscape in the low-temperature phases of the SrTiO$_{3}$.
Polarization and inversion symmetry breaking lead to asymmetries in dynamical electron scattering, and can be extracted from momentum ($\mathbf{k}$) resolved electron diffraction \cite{spence1993accurate, zuo2017advanced, deb2020imaging}.
In so-called four-dimensional scanning transmission electron microscopy (4D-STEM), an electron nanobeam (here using a probe size of $\sim1.2~nm$) is scanned across the material, and two-dimensional electron diffraction patterns, I($\mathbf{k}$), are recorded at every position (Fig. 2a) using a pixelated detector with single electron sensitivity (detector quantum efficiency $\approx$ 0.9) \cite{cowley1993configured,pennycook2015efficient,tate2016high,ophus2019four,philipp2022very}. 
In addition to intense Bragg reflections, the two-dimensional electron scattering distribution contains Kikuchi bands which arise from multiple scattering processes within the crystal lattice.
Asymmetry between Kikuchi bands related by inversion (Friedel pairs at $\pm \mathbf{k}$ momentum space coordinates) encodes polarization in the sample. 
By scanning across a sample, real space maps of polarization textures in ferroelectrics can be formed \cite{tsuda2012nanoscale, shao2017nanoscale,shao2017lattice,philipp2022very,shao2023emergent,caretta2023non}.

In SrTiO$_{3}$, the connection between polarity and asymmetry in Kikuchi band intensity is confirmed using dynamical scattering multislice simulations (see Methods).
Figures 2b and c show simulated diffraction patterns with opposite polarization added to tetragonal SrTiO$_{3}$.
The antisymmetric component in the diffraction pattern, defined as $A(\mathbf{k}) = |k^{2}|(I(\mathbf{k})-I(-\mathbf{k}))$, reveals polarization-induced intensity asymmetry in the Kikuchi bands (see Fig. S4 and note III for details).
Intensity differences between the Friedel pairs (highlighted by the black dashed box) emerge when polarization is present (right panel of Fig. 2c).
The direction of polarization is also reflected in the asymmetry: 
downward polarization leads to Kikuchi band asymmetry along the same direction. 
When the polarization is upward, the asymmetry reverses.
This intensity asymmetry is absent in the cubic, non-polar structure of SrTiO$_{3}$ (Fig. S5).

Mapping the low-temperatures phases in SrTiO$_{3}$ also requires stable cryogenic temperatures that can reach the quantum paraelectric phases \cite{zhu2021cryogenic} below $T_{AFD}$ and ideally below $T_{q}$.
Traditional liquid nitrogen cooling in STEM is limited to temperatures above 100 K \cite{kumar2024calibrating,schnitzer2025atomic}.
Liquid helium sample stages are limited by ultra-short experimental durations, sample motion, and poor variable temperature control \cite{zhu2021cryogenic,mun2024atomic}.
The recent development of a low-vibration liquid helium-cooled STEM sample stage with variable temperature control \cite{rennich2025ultra} enables access to the low-temperature polar phases of SrTiO$_{3}$.
Figure 2a shows a simplified experimental setup, which involves 4D-STEM imaging of a SrTiO$_{3}$ single crystal at cryogenic temperatures down to $\sim$ 20 K (see experimental parameters in Methods).

{\subsection*{Quantum paraelectric phases defined by nanoscale polar textures}}

Experimental maps reveal that SrTiO$_{3}$ is locally polar, hosting finite polarization below the AFD transition.
\textcolor{blue!50}{Polarization is extracted by mapping the center-of-mass (COM) shift of Kikuchi band intensity in the raw diffraction pattern, $\mathbf{\Delta}_\mathbf{k}(\mathbf{r})$, at each scan position $\mathbf{r}$ (see Figs. S6-S8 and notes IV-V for details).}
In the room temperature cubic phase, no polarization is detected (Figs. S5 and S14), as expected for the non-polar paraelectric phase.
The AFD transition below $T_{AFD}$ in SrTiO$_{3}$ also maintains global centrosymmetric symmetry. 
Upon cooling below $T_{AFD}$, however, local polar textures emerge in $\mathbf{\Delta_k}(\mathbf{r})$ over the same field of view.
Figure 2d shows a map of $\mathbf{\Delta_k}(\mathbf{r})$ in SrTiO$_{3}$ at 69 K.
The direction and color of the arrows reflect the local polar direction.
The $40 \times 80 ~nm^{2}$ field of view contains multiple small domains with opposite polarization directions.
Domains with opposite polarity (regions I and II) show a reversal in $A(\mathbf{k})$, the intensity asymmetry in the Kikuchi band (Fig. 2e, raw data in Fig. S4), in agreement with dynamical scattering multislice simulations. 
Other experimental effects of sample mis-tilt, noise, and electron beam influence on the polarization measurement are systematically evaluated (Figs. S9–S13 and notes VI-VIII).
$\mathbf{\Delta_k}(\mathbf{r})$ maps reveal that below $T_{AFD}$, polar order emerges in the form of spatially fluctuating nanoscale regions, making it locally non-centrosymmetric.

The spatially fluctuating landscape involves polar nanodomains with a characteristic length of $\sim$ 20 $nm$.
Figure 3a displays a real space map of $\mathbf{\Delta_k}(\mathbf{r})$ in SrTiO$_{3}$ at 69 K over a larger $140 \times 140 ~nm^{2}$ field of view.
The color and transparency represents the direction and amplitude of $\mathbf{\Delta_k}(\mathbf{r})$ (shown without the arrows for simpler visualization).
The map represents static configurations over the minute-timescale of image acquisition; fast temporal fluctuations are averaged over.
The real space map exhibits a highly textured arrangement of spatially fluctuating polar nanodomains, involving multiple polarization directions over tens of nanometers.
The spatial organization of polar order is then quantified through the correlation function, $C(\mathbf{r}) = \left\langle \mathbf{\Delta_k}(\mathbf{r})  \cdot \mathbf{\Delta_k}(\mathbf{r+r'}) \right\rangle_{\mathbf{r'}}$, where $\left\langle ... \right\rangle_{\mathbf{r'}}$ denotes the spatial average. 
Figure 3b shows $C(\mathbf{r})$ and the radial profile, $C(r)$.
The correlation function first exhibits a rapid exponential decay, indicative of short-range polar order. 
The short-distance decay is described by $C(r) \sim  C_{0}\exp[-r/\xi] $, where $\xi$ characterizes the domain size or correlation length.
Fitting yields $\xi \sim 20 \pm 0.6~\text{$nm$}$ at this temperature (see Methods for more details), establishing the characteristic size of the emergent polar nanodomains in SrTiO$_{3}$.

The local polar nanodomains in quantum paraelectric SrTiO$_{3}$ also develop a periodic spatial ordering.
In the correlation function, an additional peak emerges at a length scale $\lambda \approx $~25 $nm$, indicating the presence of longer-range spatial correlations beyond the typical size of individual nanodomains.
This extra peak suggests that the small polar nanodomains are not randomly distributed; 
instead, they exhibit periodic spatial order among themselves over tens of nanometers. 
Such behavior points to an unusual intermediate level of polar order below $T_{AFD}$, where nanodomains exhibit a degree of spatial modulation.
Inelastic scattering on SrTiO$_{3}$ measurements reveal a polar-acoustic anomaly, involving an incomplete softening of a transverse acoustic mode at a finite wavevector, $q$ \cite{muller1991indication,fauque2022mesoscopic,orenstein2025observation}.
A full softening would indicate a phase transition into a incommensurate modulation of atomic positions with a period $\lambda = 2\pi/q$.
However, the acoustic phonon dispersion shows only a slight and local minimum at $q$, a behavior analogous to the "roton" minimum in helium \cite{muller1991indication,fauque2022mesoscopic,orenstein2025observation}.
Images of the polar nanodomains show that the periodic length scale $\lambda$ represents spatial correlations between the emergent polar nanodomains, rather than a true incommmensurate structural modulation.

Cooling below $T_{q}\approx$~40 K reveals that the quantum paraelectric phase undergoes a transition akin to a polar \textit{glass} transition; the polar nanodomains fragment at lower temperatures. 
Typically, spatial correlations and order increase when cooling. 
However, instead of a growth or coalescence of domains at low temperature, the quantum paraelectric regime below $T_{q}$ is characterized by smaller and more disordered polar clusters.
Figure 3c shows a real space map of $\mathbf{\Delta_k}(\mathbf{r})$ at 35 K, below $T_{q}$ where SrTiO$_{3}$ enters the quantum paraelectric regime.
Domains are much smaller and more fragmented, with $C(r)$ displaying a more rapid decay, with $\xi \sim 14 \pm 0.4~\text{$nm$}$.  
The smoothly evolving background along the lateral direction arises from a small thickness gradient across the field of view (Figs. S16-S17 and notes IX).
Beyond domain size, the periodic ordering between the polar nanodomains is also suppressed.
The previously observed periodic ordering, which manifests as an extra peak in $C(r)$, weakens substantially. 
Thus, the structure in quantum paraelectric regime below $T_q$ is characterized by both smaller polar nanodomains and a loss of periodic self-organization between them.

{\subsection*{Formation, correlation, and fragmentation of polar domains} }

Tracking the temperature dependence reveals the unusual evolution of the spatially fluctuating polar landscape of quantum paraelectric SrTiO$_{3}$.
Figure 4a shows real space maps of $\mathbf{\Delta_k}(\mathbf{r})$ across temperatures; 
from 111 K to 23 K, the polar textures exhibit clear changes (complete datasets in Fig. S18). 
At 111 K (around $T_{AFD}$), tiny and weak polar nanodomains exist. 
At temperatures down to $\sim$ 69 K, the polar nanodomains grow in size, and forms spatial fluctuations of polarization over $\sim $ 20 $nm$ length scale.
Below 69 K, and especially below 40 K, quantum paraelectricity emerges as fragmented clusters of smaller and more disordered polar nanodomains. 
Further cooling to 23 K leads to even greater fragmentation, with the polar nanodomains shrinking further.

The transition to the quantum paraelectric regime also involves a loss of the periodic order between nanodomains. 
The Fourier transform of $\mathbf{\Delta_k}(\mathbf{r})$ encodes the periodic spatial ordering of polar nanodomains.
As shown in Fig. 4b, superlattice peaks (highlighted by arrows) at 69 K and 91 K reflect periodic ordering of the polar nanodomains.
Radial profiles of the peaks indicate a wavevector $q$ $\approx$ 0.04 $nm^{-1}$, consistent with the $\lambda\approx$ 25 $nm$ length scale associated with the periodic ordering of polar nanodomains found in real space.
The intensity of the superlattice peaks increases up to 69 K (Fig. 4c, data-analysis in Fig. S19). 
It then reverses and drops in the quantum paraelectric phase, indicating a loss of the periodic ordering of polar nanodomains.

The polar structure of quantum paraelectric SrTiO$_{3}$ thus exhibits a rich spatial landscape at the nanoscale.
Figure 4d summarizes the unusual evolution of the fluctuating polar ordering.
Polar nanodomains form around $T_{AFD}$ and self-organize into an emergent periodic textures, down to $\sim$ 69 K.
Below this temperature, quantum fluctuations play a more dominant role, leading to a reversal of the trend, with a decrease in the periodic correlation and the change of polar nanodomains into smaller clusters.
Upon entering the quantum regime below $T_{q}$  $\approx$ 40 K, the polar nanodomains dissolve into fragmented polar nanodomains, featuring smaller and more random domain patterns without long-range correlation.
The loss of order at lower temperature is reminiscent of re-entrant spin glass transitions \cite{dho2002reentrant} or inverse melting \cite{zhang2025inverse}.
Quantum paraelectricity in SrTiO$_{3}$ is thus characterized not by a complete absence of polarization but by the evolution of existing polar nanodomains through a re-entrant disordering at low temperatures.
In this picture, quantum fluctuations are entwined with an increase in spatial fluctuations in both the size of polar nanodomains and the periodic correlations between them.

{\subsection*{Discussion and outlook}}

SrTiO$_{3}$ exhibits elastic anomalies and unusual lattice dynamics at low temperature, first below the AFD transition and then upon entering the quantum paraelectric regime.
The low-temperature anomalies could possibly originate in AFD domain walls that become polar and mobile below $T_{q}$  \cite{wang2000electron,scott2012domain,pesquera2018glasslike}.
However, our direct real-space visualizations show that SrTiO$_{3}$ exhibits intrinsic polar order throughout the sample, rather than being confined to AFD domain walls. 
Inelastic neutron measurements find a "polar-acoustic" regime with intrinsic spatial fluctuations of polar order \cite{fauque2022mesoscopic, orenstein2025observation}.
In particular, they find a slight softening of the transverse acoustic phonon branch at finite but small wavevectors below $T_{AFD}$, followed by a loss of intensity below $T_{q}$.
The precise structure of these spatial fluctuations in SrTiO$_{3}$ remains unresolved as it could correspond to several real-space ordering patterns, including polarization density waves \cite{orenstein2025observation}, lamellar order \cite{guzman2023lamellar}, or long-wavelength transverse deformations \cite{fauque2022mesoscopic}.
Further, the nature of the transition below $T_{q}$, defined by a loss of the transverse acoustic branch, has likewise remained unclear.
Direct imaging reveals that SrTiO$_{3}$ below $T_{AFD}$ forms polar nanodomains that organize into a modulated structure over tens of nanometer length scales.
The nanodomains then fragment into small, uncorrelated polar textures in the quantum paraelectric regime.

The observed real-space polar textures and their evolution are reminiscent of the theoretical proposal of melted \textit{lamellar} order \cite{guzman2023lamellar}.
The coupling of polarization to a secondary mode (such as strain gradient or flexoelectricity) could allow for instabilities at a finite wavevector, giving rise to incommensurate periodic ordering of polarization, a state dubbed the lamellar phase.
Crystalline anisotropy further helps stabilize this modulation into long-range unidirectional stripes at low temperatures.
However, the lamellar phase is easily suppressed by thermal fluctuations even at low temperatures, leaving only traces of a melted modulated order defined by decaying oscillatory correlations \cite{guzman2023lamellar}.
Our visualizations of the real space textures and their spatial correlations provide a more concrete picture of the nature of the polar-modulated lamellar state.

The spatial ordering and disordering of polar nanodomains raise additional questions on the mechanisms through which external tuning parameters induce novel orders in quantum paraelectrics \cite{mitsui1961dielectric, itoh1999ferroelectricity, haeni2004room, aschauer2014competition, lee2015emergence,xu2020strain, salmani2020order,li2019terahertz,nova2019metastable,basini2024terahertz, fechner2024quenched}.
Possibly, external stimuli such as mid-infrared pulses or strain cause a long-range ferroelectric state through emergent correlations between the preformed polar nanodomains, similar to order-disorder transitions in other ferroelectrics \cite{lines2001principles, zhang2025real}.
This should also motivate the exploration of the link between unconventional superconductivity in doped SrTiO$_{3}$ \cite{schooley1964superconductivity, koonce1967superconducting,
collignon2019metallicity} and the spatially fluctuating polar state in the quantum paraelectric regime. 
The evolution of polar nanodomains as a function of electron doping could reveal the nature of the parent fluctuating polar phase from which superconductivity and quantum criticality emerge.

{\section*{Materials}}

{\subsubsection*{Sample preparation}}
Commercial SrTiO$_{3}$ single crystals (MTI corporation) were used for this study.
The SrTiO$_{3}$ lamella was prepared using standard gallium focused ion beam (FEI Helios 660) lift-out and thinning.
The samples were thinned down using an accelerating ion voltage of 30 kV with a decreasing current from 100 pA to 40 pA, and then with a fine polishing process using an accelerating voltage of 5 kV and 2 kV, and a current of 41 pA and 23 pA.

The STEM lamella sample is supported on a Cu post of the FIB grid on only one side, making the sample free to naturally contract with temperature (Fig. S1). 
This mounting geometry ensures minimal strain imparted to the lamella while maintaining thermal coupling for stable cryogenic experiment.
The thickness of the sample within the imaging region is estimated using the zero-loss peak in electron energy loss spectroscopy data (Fig. S17).
Thickness varies from $\sim53$ to $69~nm$ across the relatively large $140\times140~nm^{2}$ field of view.

{\subsubsection*{Liquid helium cryogenic electron microscopy}}
Cryogenic STEM experiments were performed using an h-Bar Instruments ULT holder which uses continuous flow through a heat exchanger that is mechanically decoupled from the sample to reduce vibrations \cite{rennich2025ultra, sung2025liquid}. 
Variable temperatures are enabled by an integrated heater and PID control. 
The sample drift rate at low temperature is measured as 0.37 \AA/s at base temperature through image tracking, lower than the range (0.6 -1.0 \AA/s) for collecting high-resolution image in conventional side-entry liquid nitrogen cryogenic holders \cite{el2018nature, goodge2020atomic}. 
For each temperature step during heating, we realign the sample location to the same region after waiting for temperature to stabilize and sample drift to become negligible. 
A complete 4D-STEM dataset is acquired in $\sim$ 2 mins, so drift during acquisition produces at most $\sim$ 4 $nm$, far smaller than the observed domain size scale.

{\subsubsection*{Four dimensional scanning transmission electron microscopy}}
Recent advances in pixelated detectors, particularly direct electron detectors with high dynamic range \cite{tate2016high,philipp2022very}, have enabled the simultaneous capture of both intense Bragg reflections and more subtle features such as Kikuchi bands (more than $10^4$ weaker than Bragg spots) which arise from multiple scattering processes within the crystal lattice.
4D-STEM measurements were performed in an aberration-corrected scanning transmission electron microscope (Thermo Fisher Scientific Spectra 300) operated at 300 kV. 
4D STEM datasets were collected by a first-generation electron microscopy pixel array detector (EMPAD), using a 1 mrad convergence semi-angle.
The probe current was 30 pA.
The real-space sampling was $140 \times140$ $nm^{2}$ for each dataset, with $\sim$ 0.69 $nm$ pixel size.
The reciprocal space area spans $14 \times14 \ $\r{A}$^{-2}$, with a $\sim$ 0.1 \r{A}$^{-1}$ pixel size.
Each diffraction pattern was acquired with 1 ms dwell time, with an additional 0.86 ms for readout time.

{\subsubsection*{Multislice simulation}}
Dynamical scattering multislice simulations were carried out using abTEM \cite{madsen2021abtem} and custom scripting Python code. 
The slice thickness was set to 0.5 $nm$.
The frozen phonon method was used to account for thermal diffuse scattering, with a total of 30 configurations per slice. 
The convergence semi-angle is 1 mrad, similar to the experimental conditions.
The acceleration voltage was set to 300 kV and defocus to 0 $nm$.
For the nanobeam diffraction patterns shown in Figs. S4, S5, S7 and S10, a $8\times8~nm^{2}$ cell with $60~nm$ thickness (close to sample thickness) was constructed.
The B-factor ($\sqrt{u^2}$) in the frozen phonon is set to 0.1 and 0.12 \text{\AA} for tetragonal and cubic SrTiO$_{3}$ to mimic the temperature effect on diffuse scattering.
For the simulations in Fig. S16, a series of thicknesses ranging from $5-70~nm$ along the beam direction was used.

{\subsubsection*{Autocorrelation function}}
To characterize the size and periodicity of polar domains, we computed the 2D autocorrelation function of $\mathbf{\Delta}_\mathbf{k}(\mathbf{r})$, derived from a complex-valued 2D field $\Delta(\mathbf{r})\exp(i\varphi(\mathbf{r}))$, where $\Delta(\mathbf{r})$ and $\varphi(\mathbf{r})$ denote the amplitude and angle of $\mathbf{\Delta}_\mathbf{k}(\mathbf{r})$, respectively. 
The autocorrelation function was then obtained by computing the inverse Fourier transform of the summed power spectra of each component:
\begin{equation}
C(\mathbf{r}) = \mathcal{F}^{-1}\left[|\mathcal{F}[\Delta(\mathbf{r})\exp(i\varphi(\mathbf{r}))]|^2 \right],
\end{equation}
where $\mathcal{F}$ and $\mathcal{F}^{-1}$ denote the forward and inverse 2D Fourier transforms, respectively. 

The 2D autocorrelation map (inset of Figs. 3b, d) was then radially averaged to obtain the radial profile, $C(r)$, as a function of radial distance $r$ (Figs. 3b, d).
$C(r)$ was used for quantification of the correlation length, ($\xi$), a measure of the characteristic domain size and short range order. 
Since the autocorrelation function follows exponential decay, we fit it using the standard form 
\begin{equation}
y(r) = \Delta \exp\left(-\frac{r}{\xi}\right).
\end{equation}

To obtain $\xi$, we converted the radial profile to a logarithmic scale and then performed linear fitting using the transformed expression to get $\xi$.
\begin{equation}
log(y) = \log(\Delta) - \frac{r}{\xi} 
\end{equation}


\clearpage 

%
\bibliography{references} 
\bibliographystyle{sciencemag}

%
%
%
%
%
%


\paragraph*{Acknowledgements:}
We acknowledge insightful discussions with John Heron, Haozhi Sha, Yu-Tsun Shao and Haoyang Ni.
Y. Z., S. H. S. and I. E. were supported by the Rowland Institute at Harvard.
R. H. and N. A. were supported by U.S. Department of Energy, Basic Energy Sciences, under award DE-SC0024147.
C. L. and P. Y. were financially supported by the National Natural Science Foundation of China (NSFC Grant Nos. 52025024, 52388201 and 12421004), and the National Key R\&D program of China (Grant No. 2023YFA1406400).
Liquid helium cryogenic electron microscopy experiments were performed at the Michigan Center for Materials Characterization ((MC)$^{2}$) at the University of Michigan.
Focused ion beam sample preparation was performed at the Harvard University Center for Nanoscale Systems (CNS); a member of the National Nanotechnology Coordinated Infrastructure Network (NNCI), which is supported by the National Science Foundation under NSF award no. ECCS-2025158.

\paragraph*{Author contributions:}
Y. Z. and I. E. conceived the study. 
Y. Z., S. H. S., N. A., M. G., R. H., and I. E. performed microscopy experiments.
C. L. and P. Y. helped prepare samples.
Y. Z. S. H. S., R. H. and I. E. analyzed data and wrote the manuscript with contributions from all authors.
\paragraph*{Competing interests:}
I. E. and R. H. are inventors on a patent related to the liquid helium sample holder invention. 
I. E., R. H. and M. G. have have a financial interest in h-Bar Instruments which is commercializing the invention. Other authors declare no competing interests.
\textcolor{blue!50}{\paragraph*{Data availability:}
Electron microscopy data corresponding to the figures in the manuscript are available at Zenodo https://doi.org/10.5281/zenodo.20300700. 
Additional data are available from the corresponding authors upon request.}
\textcolor{blue!50}{\paragraph*{Code availability:}
Analysis codes for the center-of-mass measurement and spatial map are available at Zenodo https://doi.org/10.5281/zenodo.20300700.}


\paragraph*{Supplementary materials:}
Supplementary materials incliudes supplementary notes I-IX, figures S1-S19 and references \textit{(68-\arabic{enumiv})}\\ 


\newpage

\begin{figure}
    \centering
    \includegraphics[width=\textwidth]{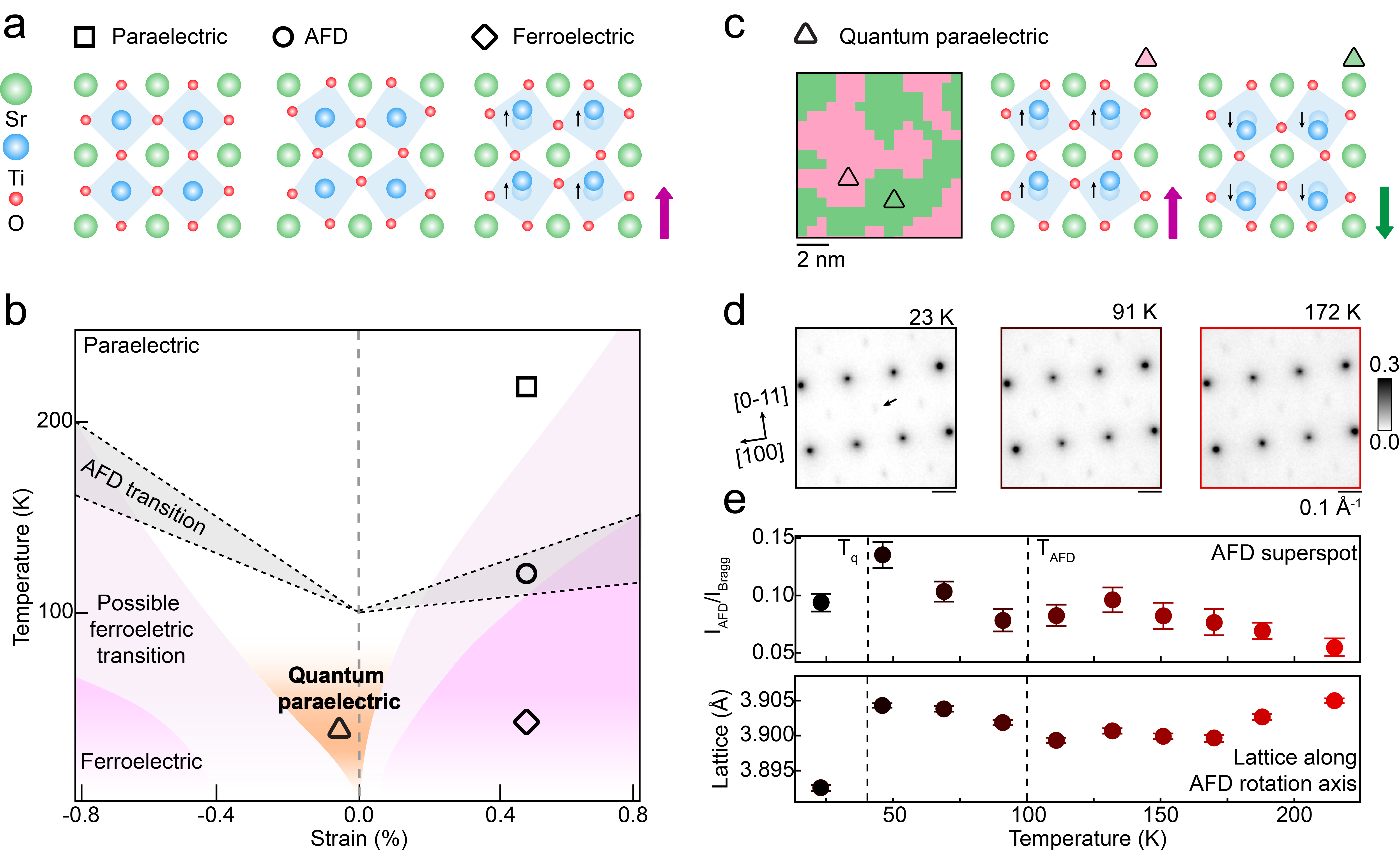}
    \caption{\textbf{Structural transitions in the SrTiO$_{3}$ sample.}
    (a) Structural phases of bulk SrTiO$_{3}$.
    From the cubic paraelectric to the antiferrodistortive (AFD) tetragonal phase, TiO$_{6}$ octahedra undergo out-of-phase rotations, lowering the symmetry to tetragonal.
    In the ferroelectric phase, titanium atoms shift away from the center of the TiO$_{6}$ octahedron, leading to an electric dipole moment.
    (b) Phase diagram of SrTiO$_{3}$ as a function of temperature and strain. 
    Stabilization of a long-range ferroelectric phase can also be realized using calcium doping or oxygen isotope substitution.
    The parent SrTiO$_{3}$ phase transition follow the gray dashed line,
    entering the AFD phase around T$_{AFD} \sim$ 105 K followed by a quantum paraelectric regime at lower temperatures.
    The phase diagram is adapted from \cite{li2006phase}.
    (c) Schematic of polar nanodomains in quantum paraelectric SrTiO$_{3}$, with spatial fluctuations of polarization orientations at the nanoscale (field of view 10 $\times$ 10 $nm^{2}$).
    (d) Electron diffraction patterns collected at different temperatures. 
    Black arrows mark the AFD superspots located at half-integer positions. 
    (e) Quantification of (upper) AFD superspot intensity, $I_{AFD}/I_{Bragg}$ , and (lower) lattice parameter along the AFD-rotation axis. 
    Two anomalies near  
    $T_{AFD}$  and $T_{q}$ are highlighted by black dashed lines. 
    The SrTiO$_{3}$ sample used for STEM imaging thus retains the bulk-like structural transitions.
    }
    \label{fig1}
\end{figure}

\begin{figure}
    \centering
    \includegraphics[width=\textwidth]{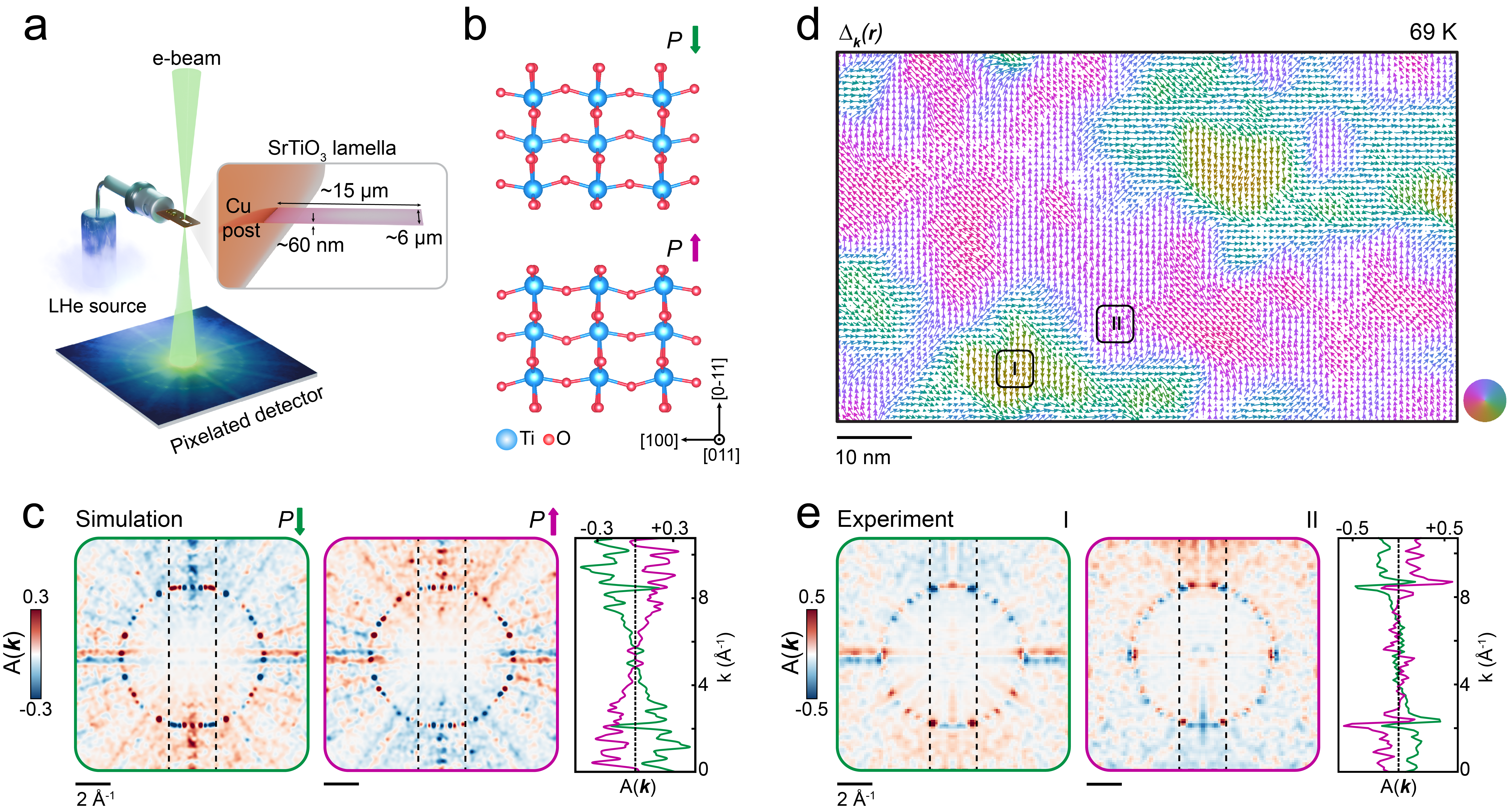}
    \caption{\textbf{Imaging polar textures in cryogenic SrTiO$_{3}$.}
    (a) Experimental setup for liquid helium cryogenic scanning transmission electron microscopy (STEM) with 
    nanobeam four-dimensional STEM (4D-STEM) for mapping polarization in SrTiO$_{3}$.
    The inset illustrates the geometry of SrTiO$_{3}$ lamellar sample prepared by focused ion beam.
    (b) Structures of tetragonal SrTiO$_{3}$ with polarization along [$00\bar{1}$] (down) and [$001$] (up) directions, viewed along [$011$].
    For simplicity, only titanium and oxygen atoms are shown.   
    (c) Simulated nanobeam electron diffraction based on two opposite polarization states.
    The antisymmetric component in the diffraction pattern, defined as $A(\mathbf{k}) = |k^{2}|(I(\mathbf{k})-I(-\mathbf{k}))$, reveals polarization-induced intensity asymmetry in the Kikuchi bands.
    Asymmetry along the $\mathbf{k}_{y}$ direction is apparent and switches between $P$-up and $P$-down structures. 
    The black dashed box is selected to extract intensity profiles (right panel).
    (d) Real-space map of center-of-mass shifts of Kikuchi bands, $\mathbf{\Delta_{k}}(\mathbf{r})$. 
    The field of view is 40 $\times$ 80 $nm^{2}$ and the temperature is 69 K.
    Arrows reflect the amplitude and the color represents the direction.
    (e) Averaged diffraction patterns from two boxed regions (I and II) in (d).
    Consistent with simulations, $A(\mathbf{k})$ reveals polarization-induced intensity asymmetry in the diffraction pattern.
    The right panel shows line profiles of $A(\mathbf{k})$ for regions of opposite polarization.
    }
    \label{fig2}
\end{figure}

\begin{figure}
    \centering
    \includegraphics[width=\textwidth]{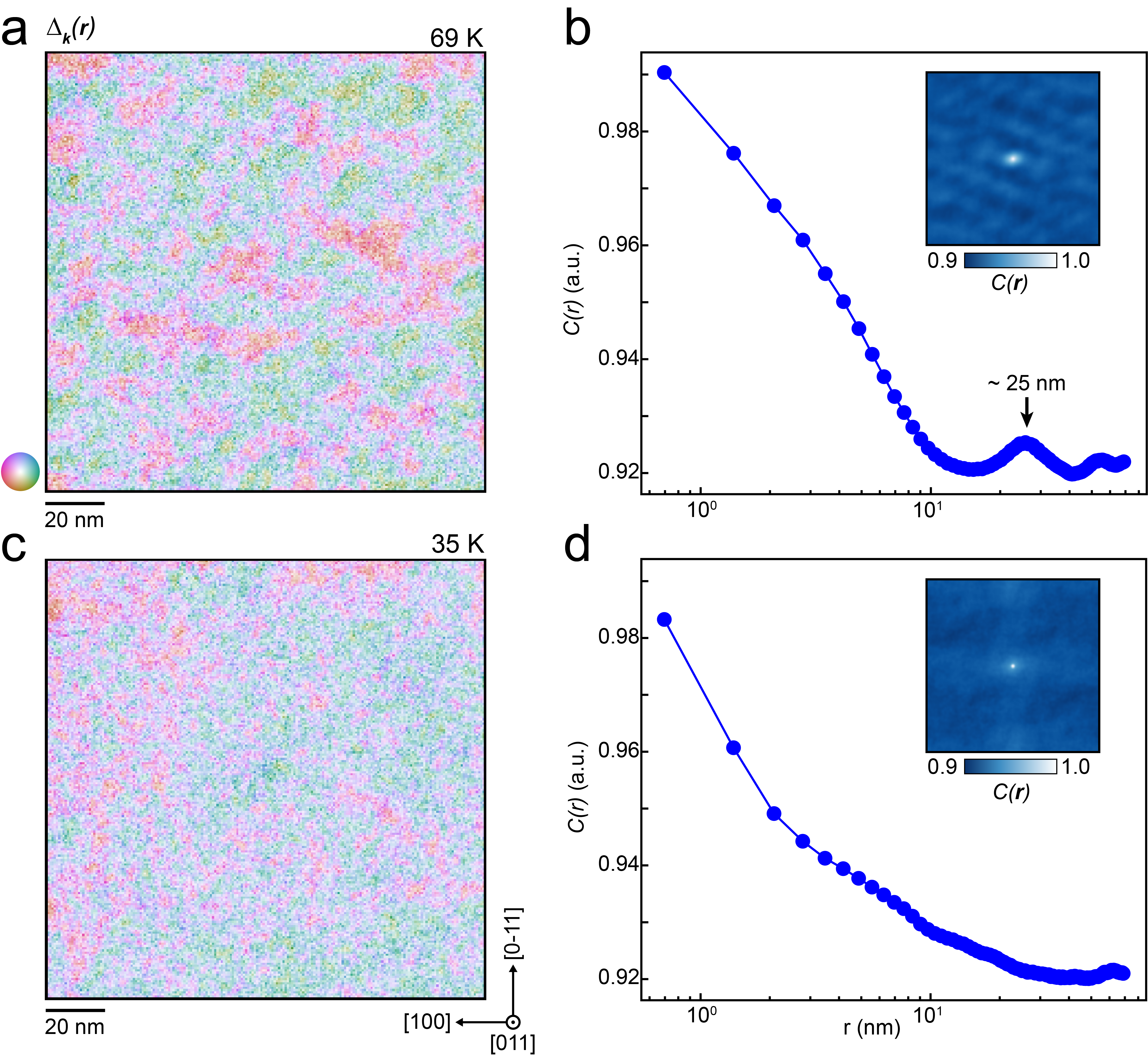}
    \caption{\textbf{Nanoscale polar textures in quantum paraelectric SrTiO$_{3}$.}
    (a, c) Large field-of view map of $\mathbf{\Delta}_\mathbf{k}(\mathbf{r})$ at 69 K and 35 K, respectively.
    The field of view is $140\times140$ $nm^{2}$.
    The color and transparency represents the direction and amplitude of $\mathbf{\Delta}_\mathbf{k}(\mathbf{r})$.
    (b, d) The inset is a 2D map of autocorrelation function of $\mathbf{\Delta}_\mathbf{k}(\mathbf{r})$, $C(\mathbf{r})$. The plot shows the radial profile, $C(r)$.
    }
    \label{fig3}
\end{figure}

\begin{figure}
    \centering
    \includegraphics[width=\textwidth]{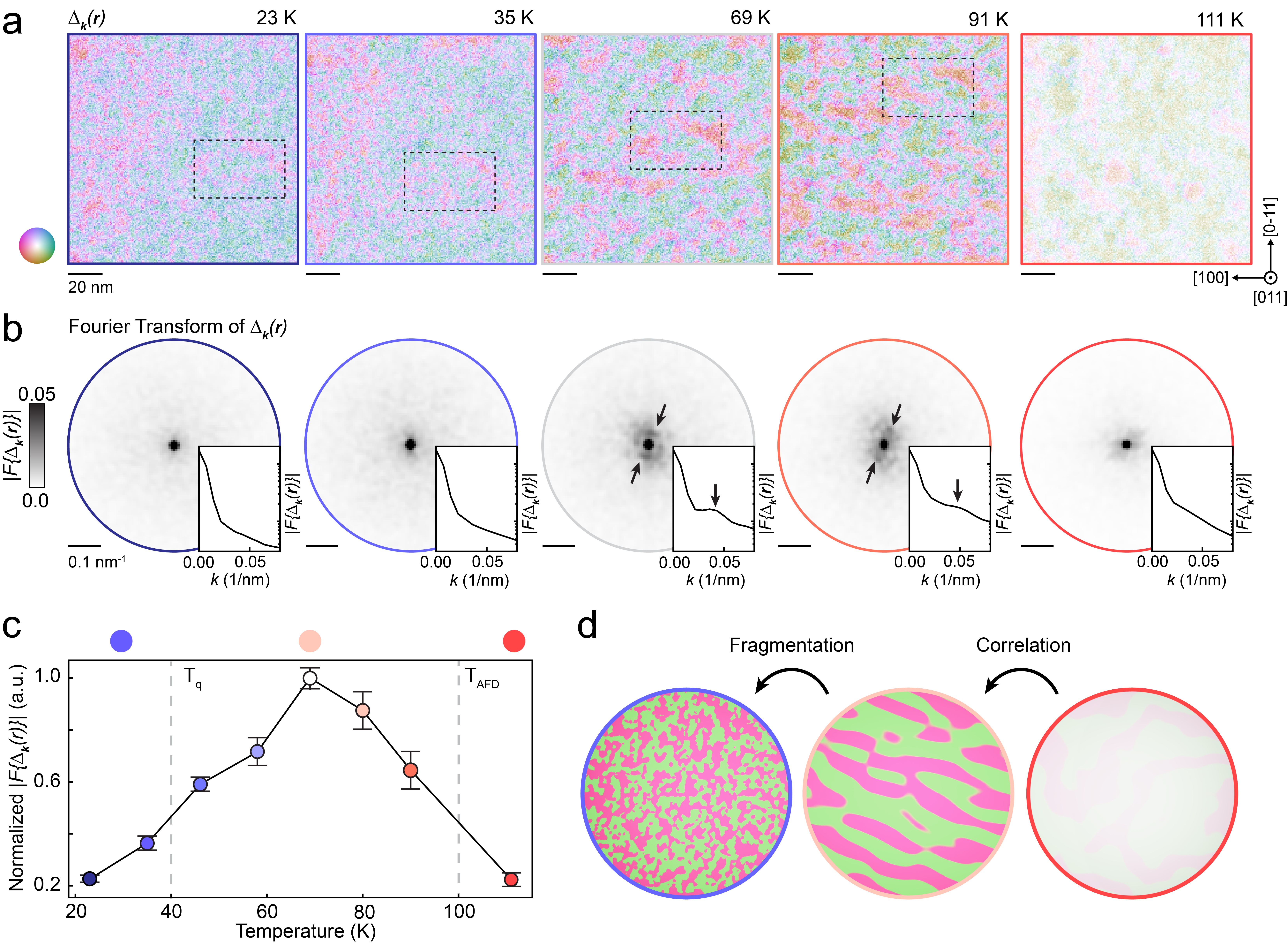}
    \caption{\textbf{Evolution of a spatially fluctuating polar landscape at low temperature.}
    (a) Large field-of view map of $\mathbf{\Delta}_\mathbf{k}(\mathbf{r})$ in SrTiO$_{3}$ at 23 K, 35 K, 69 K, 91 K and 111 K.
    The field of view is $140\times140$ $nm^{2}$.
    The color and transparency represents the direction and amplitude of $\mathbf{\Delta}_\mathbf{k}(\mathbf{r})$.
    Dashed rectangles denote matching features across temperatures over roughly the same field of view.
    (b) Fourier transforms of $\mathbf{\Delta}_\mathbf{k}(\mathbf{r})$ maps.
    The inset is the radial profile of the Fourier transform.
    Arrow highlights a superlattice peak ($q \approx0.04~nm^{-1}$, corresponding to a wavenlength $\lambda \approx 25~nm$ in real space) at intermediate temperatures.
    (c) Superlattice peak intensity as a function of temperature (raw data and analysis in Figs. S18-S19).
    (d) Schematic of the polar landscape in quantum paraelectric SrTiO$_{3}$ based on our real-space observations.
    Polar nanodomains form below $T_{AFD} \approx$ 105 K and develop periodic spatial correlations.
    The nanodomains fragment when SrTiO$_{3}$ enters the quantum paraelectric regime below $T_{q}$ $\approx$ 40 K.
    }  
    \label{fig4}
\end{figure}

\clearpage




\end{document}